\documentclass[a4paper,twoside,12pt]{article}
\input{obsjkt.sty}

\newcommand{\Msun}{\ensuremath{~{\rm M}_\odot}}                   
\newcommand{\Rsun}{\ensuremath{~{\rm R}_\odot}}                   
\newcommand{\rhosun}{\ensuremath{~\rho_\odot}}                    
\newcommand{\Teff}{\ensuremath{T_{\rm eff}}}                      
\newcommand{\degr}{\ensuremath{^\circ}}                           
\renewcommand{\kms}{~km~s$^{-1}$}                                 
\newcommand{\chisq}{\ensuremath{\chi^{\,2}}}                      
\newcommand{\chir}{\ensuremath{\chi_\nu^{\,2}}}                   

\newcommand{\etal}{\textit{et al.}}                               
\newcommand{\corot}{\textit{CoRoT}}
\newcommand{\kepler}{\textit{Kepler}}
\newcommand{\tess}{\textit{TESS}}

\newcommand{\gaia}{\textit{Gaia}}

\newcommand{\debcat}{\textit{DEBCat}}
\newcommand{\targ}{KIC 5359678}

\newcommand{\Msunnom}{\hbox{$\mathcal{M}^{\rm N}_\odot$}}
\newcommand{\Rsunnom}{\hbox{$\mathcal{R}^{\rm N}_\odot$}}
\newcommand{\Lsunnom}{\hbox{$\mathcal{L}^{\rm N}_\odot$}}

\usepackage{rotating}

\begin{document} 

\OBSheader{Rediscussion of eclipsing binaries: \targ}{J.\ Southworth}{2022 June}

\OBStitle{Rediscussion of eclipsing binaries. Paper IX. \\ The solar-type system \targ}

\OBSauth{John Southworth}

\OBSinstone{Astrophysics Group, Keele University, Staffordshire, ST5 5BG, UK}


\OBSabstract{\targ\ is a 6.231-d period F-type eclipsing binary system whose component stars both show starspot activity. It was observed by the \kepler\ satellite in long cadence for the full four-year duration of the mission. Wang \etal\ \cite{Wang+21mn} obtained radial velocity measurements of the two stars and analysed these plus the \kepler\ data to study their spot activity and measure their physical properties, but left several questions unanswered. We have performed an independent analysis and determined the masses ($1.252 \pm 0.018$ and $1.065 \pm 0.013$\Msun) and radii ($1.449 \pm 0.012$ and $1.048 \pm 0.017$\Rsun) of the stars to high precision. The distance we find to the system is slightly shorter than that from \gaia\ EDR3 for unknown reason(s). We also investigated the precision of the numerical integration applied to the model light curve to match the 1765-s sampling cadence of the \kepler\ observations. We found that ignoring this temporal smearing leads to biased radius measurements for the stars: that for the primary is too small by 4$\sigma$ and that for the secondary is too large by 10$\sigma$. Doubling the sampling rate of the model light curve is sufficient to remove most of this bias, but for precise results a minimum of five samples per observed datapoint is required.}


\section*{Introduction}

Detached eclipsing binaries (dEBs) are crucial to stellar physics because the properties of their component stars can be measured directly from observed light and radial velocity (RV) curves \cite{Andersen91aarv,Torres++10aarv}. These direct measurements can then be used to constrain and to calibrate theoretical models of stellar evolution \cite{ClaretTorres18apj,Valle+18aa,Tkachenko+20aa}. Many thousands of dEBs are now known \cite{Watson++06sass,Samus+17arep,Jayasinghe+18mn,Thiemann+21mn}. Particularly important contributions to the numbers of dEBs for which good light curves are available have recently been made by space missions such as \corot\ \cite{Deleuil+18aa}, \kepler\ \cite{Kirk+16aj} and \tess\ \cite{JustesenAlbrecht21apj,IJspeert+21aa,Prsa+21apjs}. An extensive review of the impact of space photometry on binary star science can be found in Southworth \cite{Me21univ}.

\targ\ was found to be a dEB using data from the \kepler\ satellite \cite{Kirk+16aj,Prsa+11aj,Slawson+11aj,Debosscher+11aa} with a morphology value of 0.27 which indicates that it is well-detached \cite{Matijevic+12aj}. Armstrong \etal\ \cite{Armstrong+14mn} determined effective temperature (\Teff) values for the two components of $6713 \pm 405$~K and $6237 \pm 623$~K. Qian \etal\ \cite{Qian+18apjs} determined the \Teff\ of the system to be $6510 \pm 70$~K and its spectral type to be F5 from a medium-resolution ($R=1800$) spectrum obtained using the LAMOST spectroscopic telescope \cite{Cui+12raa} survey of the \kepler\ field \cite{Decat+16apjs}.

Wang \etal\ \cite{Wang+21mn} (hereafter W21) presented a detailed study of \targ\ which concentrated on the characteristics of the starspots on the stellar surfaces. To determine the physical properties of the stars, W21 modelled the light curve from the \kepler\ satellite together with RVs from a set of 58 medium-resolution ($R=7500$) LAMOST spectra \cite{Zong+20apjs} using the {\sc phoebe2} code \cite{Prsa+16apjs}. They gave two sets of physical properties for the system, for eccentric and circular orbits, which are formally identical but with errorbars differing by as much as a factor of three. They also quoted no errorbar for the mass of the secondary star, and did not mention whether they accounted for the cadence of the observations obtained by the \kepler\ satellite. For these reasons, and to see if the properties of the system can be established to a precision of 2\% or better \cite{Andersen91aarv,Me20obs}, we present below a reanalysis of \targ. Basic information on the system is summarised in Table~\ref{tab:info}.

\begin{table}[t]
\caption{\em Basic information on \targ. \label{tab:info}}
\centering
\begin{tabular}{lll}
{\em Property}                            & {\em Value}                 & {\em Reference}                   \\[3pt]
\kepler\ Input Catalog designation        & \targ                       & \cite{Brown+11aj}                 \\
\kepler\ Object of Interest designation   & KOI 6569                    & \cite{Borucki+11apj} and updates  \\
\textit{Gaia} EDR3 designation            & 2101510803402761344         & \cite{Gaia21aa}                   \\
\textit{Gaia} EDR3 parallax               & $0.6151 \pm 0.0135$ mas     & \cite{Gaia21aa}                   \\
$B$ magnitude                             & $14.905 \pm 0.023$          & \cite{Henden+15aas}               \\
$V$ magnitude                             & $14.209 \pm 0.056$          & \cite{Henden+15aas}               \\
$H$ magnitude                             & $12.927 \pm 0.029$          & \cite{Cutri+03book}               \\
$K_s$ magnitude                           & $12.862 \pm 0.030$          & \cite{Cutri+03book}               \\
Spectral type                             & F5                          & \cite{Wang+21mn}                  \\[3pt]
\end{tabular}
\end{table}


\section*{Observational material}

The \kepler\ satellite is a 0.95~m Schmidt reflecting telescope with a focal plane contining 42 CCDs, launched by NASA in March 2009 and placed into an Earth-trailing heliocentric orbit \cite{Borucki+10sci,Borucki16rpph}. Its mission was to obtain photometric observations of a single patch of sky (the \kepler\ Field) for four years in order to find transits of extrasolar planets. Due to a limit on the amount of data that could be returned to Earth, \kepler\ could only be used to observe 170\,000 objects simultaneously. Consecutive observations were summed into single ``long cadence'' observations with an effective duration of 1765.5~s, and the count rates of pixels in the region of selected targets were transmitted to Earth. A subset of 512 of these targets could be observed in ``short cadence'', where individual observations were summed into data with an effective duration of 58.8~s. \kepler\ observations were divided into quarters (three months) due to the rotation of the spacecraft around its optical axis to keep its solar panels illuminated by the Sun. These data were reduced by the \kepler\ mission Science Operations Center (SOC) \cite{Jenkins+10apj2}.

\begin{figure}[t] \centering \includegraphics[width=\textwidth]{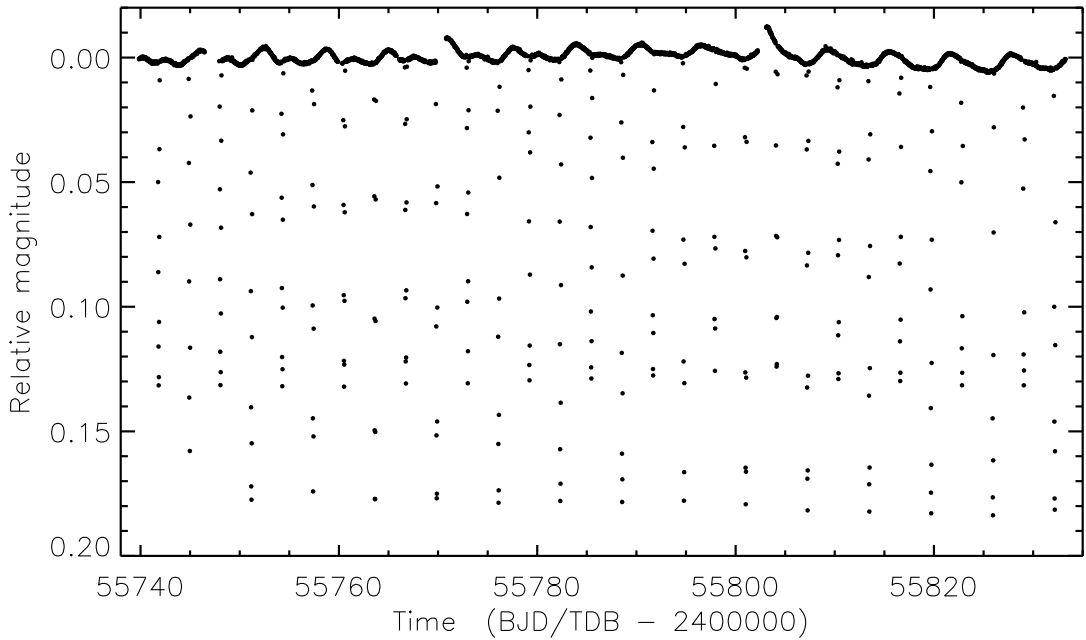} \\
\caption{\label{fig:time} \kepler\ quarter 10 light curve of \targ. The spot activity
can be seen as continual variations in the data outside eclipse, and two jumps in
the data are also visible after short interruptions to the observations.} \end{figure}

\targ\ was observed for the entirety of the \kepler\ mission, from quarters Q0 to Q17, in long cadence. The first datapoint was taken on 2009/05/13 and the last on 2013/05/11. Small gaps occured within this time interval when \kepler\ paused observations for technical reasons or to transmit data to Earth. The data were downloaded from the MAST archive\footnote{Mikulski Archive for Space Telescopes, \\ \texttt{https://mast.stsci.edu/portal/Mashup/Clients/Mast/Portal.html}} and converted to relative magnitude. Rejection of unreliable data (typically represented by ``Inf'' or ``Nan'') left 64\,073 measurements. We chose to work with the standard aperture photometry (SAP) rather than the pre-search data conditioning (PDC) data, after verifying that the differences were negligible for our purposes. An example light curve, chosen at random to be from Q10, is shown in Fig.\,\ref{fig:time}.


\section*{Analysis of the \kepler\ light curve}

\begin{figure}[t] \centering \includegraphics[width=\textwidth]{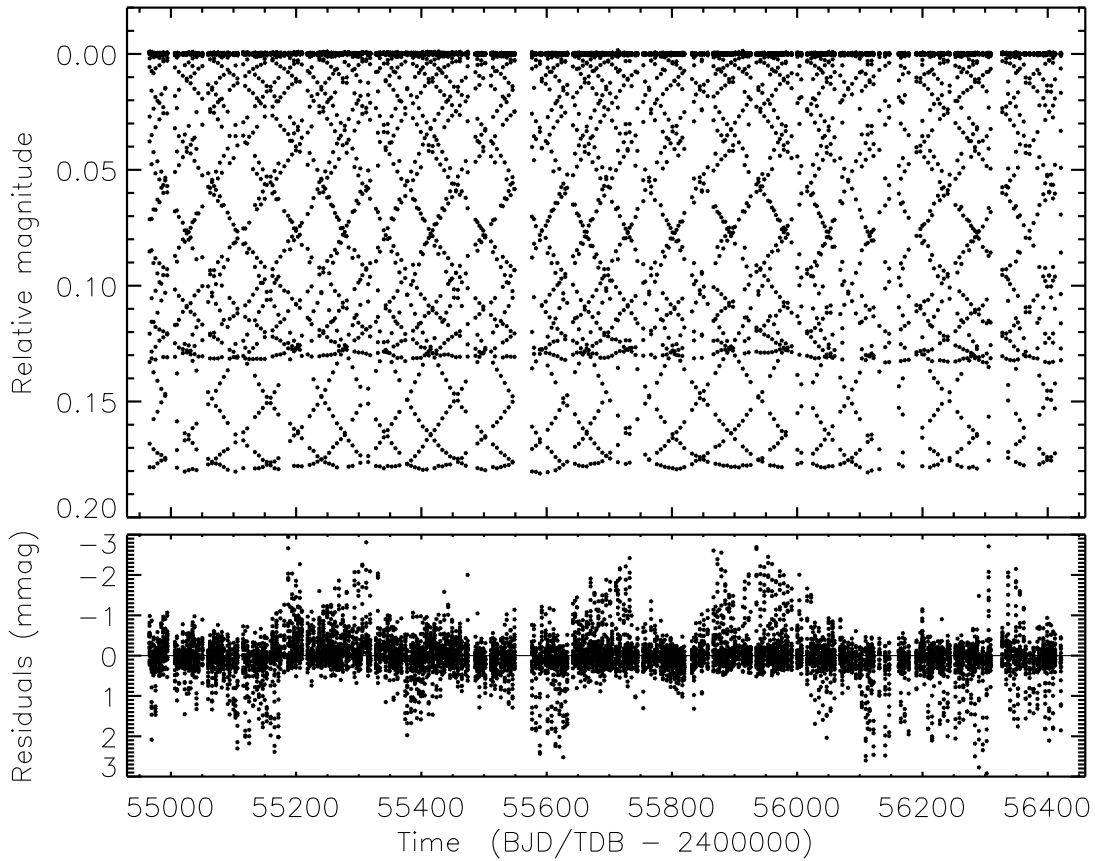} \\
\caption{\label{fig:fit} The \kepler\ light curve of \targ\ as analysed (top panel)
with the residuals of the best fit for a circular orbit (lower panel).} \end{figure}

The majority of the photometric observations occur outside eclipse, so hold negligible information on the physical properties of the stars. The starspot variation also affects the brightness of the system during eclipse but occurs on much longer timescales. We tackled both problems by extracting the data around each eclipse from the overall light curve, fitting a polynomial to the out-of-eclipse data, and subtracting the polynomial (in magnitude units) to remove the slow variations due to starspots and instrumental effects. All eclipses were manually inspected and those with insufficient data either before or after, or with more than two datapoints missing, were rejected. This left a total of 8106 observations with the main effects of spot activity removed (Fig.~\ref{fig:fit}).

We then proceeded to analyse the data with version 41 of the {\sc jktebop}\footnote{\texttt{http://www.astro.keele.ac.uk/jkt/codes/jktebop.html}} code \cite{Me++04mn2,Me13aa}. Fitted parameters included the sum and ratio of the fractional radii ($r_{\rm A}+r_{\rm B}$ and $k = \frac{r_{\rm B}}{r_{\rm A}}$ where $r_{\rm A} = \frac{R_{\rm A}}{a}$, $r_{\rm B} = \frac{R_{\rm B}}{a}$, $R_{\rm A}$ and $R_{\rm B}$ are the radii of the stars, and $a$ is the semimajor axis of the relative orbit), the orbital inclination ($i$) and period ($P$), a reference time of primary minimum ($T_0$; when star~A is eclipsed by star~B), and the central surface brightness ratio of the two stars ($J$).

For limb darkening (LD) we adopted the quadratic law, fitted the linear coefficient for each star ($u_{\rm A}$ and $u_{\rm B}$) and fixed the quadratic coefficients ($v_{\rm A}$ and $v_{\rm B}$) to theoeretical values from Sing \cite{Sing10aa}. We found third light to be very small so fixed it at zero. One important consideration is the long sampling cadence -- one observation every 1765~s -- and we accounted for that by numerically integrating the model light curve to match the observed one \cite{Me11mn}.

\begin{figure}[t] \centering \includegraphics[width=\textwidth]{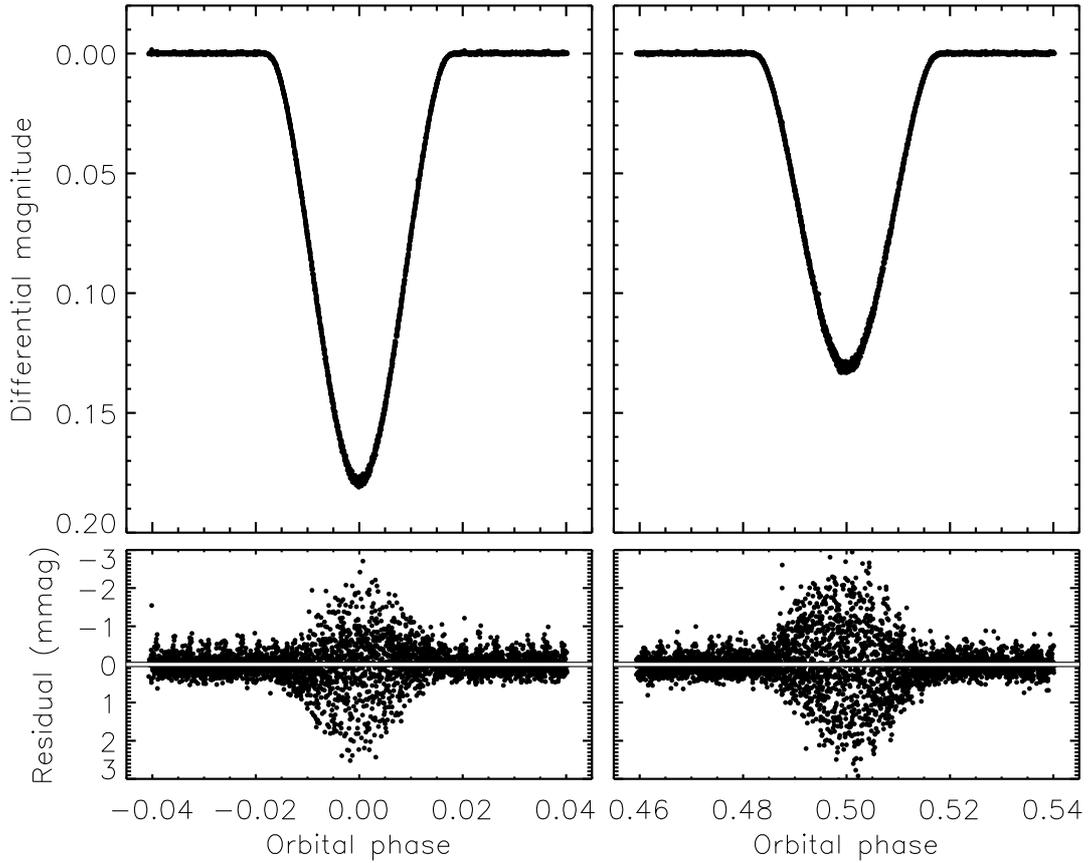} \\
\caption{\label{fig:ecl} The \kepler\ light curve of \targ\ (filled circles) around the primary (left) and
secondary (right) eclipses. The best fit is not plotted as it is indistinguishable from the data. The lower
panels show the residuals of the fit with the line of zero residual overplotted in white.} \end{figure}

We obtained solutions for a circular and an eccentric orbit. In the latter case we fitted for $e\cos\omega$ and $e\sin\omega$ where $e$ is the orbital eccentricity and $\omega$ is the argument of periastron. In both cases we included the RV measurements of the two stars published by W21 in order to measure the masses and radii of the stars. The errorbars on the three datasets (light curve, RVs of star~A, RVs of star~B) were each scaled to give a reduced \chisq\ of $\chir=1$ for that dataset.

The best fit to the light curve for the circular orbit can been seen in Fig.~\ref{fig:fit}. The residuals in this figure show systematic trends with time which occur because our approach to dealing with the spot activity does not account for the partial obscuration of spots during eclipses. Our best fit therefore systematically over- or under-predicts the eclipse depth in a way which changes on a characteristic timescale of approximately 200 days. This is much longer than the rotation period or the timescale over which individual starspots appear and disappear (see Fig.\,\ref{fig:time}), but is much shorter than the time interval covered by the data.

Another visualisation of the situation is shown in Fig.~\ref{fig:ecl}, where the data have been converted into orbital phase and shown in close-up around the eclipses. The best fit is clearly a good representation of the data, but the residuals significantly increase during both eclipses. As changes in eclipse depth are driven primarily by spots on the surface of the eclipsed star, this tells us that both stars show spot activity and that their activity levels are comparable.

\begin{table} \centering
\caption{\em \label{tab:jktebop} Parameters of the {\sc jktebop} best fits to the
\kepler\ light curve and published RVs of \targ. The uncertainties are 1$\sigma$ and
are the larger of the Monte Carlo and residual-permutation options for each parameter.}
\setlength{\tabcolsep}{4pt}
\begin{tabular}{@{}l@{\hspace*{-3pt}}cc}
{\em Parameter}                           &       {\em Circular orbit}        &        {\em Eccentric orbit}       \\[3pt]
{\it Fitted parameters:} \\
Primary eclipse (BJD/TDB)                 &   2455402.26675$\pm$0.00002       &   2455402.26670$\pm$0.00002        \\
Orbital period (d)                        &    6.23060994$\pm$0.00000024      &    6.23060991$\pm$0.00000028       \\
Orbital inclination (\degr)               & $      85.537   \pm  0.042      $ & $      85.527   \pm  0.049      $  \\
Sum of the fractional radii               & $       0.13246 \pm  0.00044    $ & $       0.13255 \pm  0.00049    $  \\
Ratio of the radii                        & $       0.724   \pm  0.022      $ & $       0.732   \pm  0.027      $  \\
Central surface brightness ratio          & $       0.7324  \pm  0.0015     $ & $       0.7253  \pm  0.0129     $  \\
Third light                               &              0.0  (fixed)         &              0.0  (fixed)          \\
Linear LD coefficient star A              & $       0.293   \pm  0.008      $ & $       0.297   \pm  0.011      $  \\
Linear LD coefficient star B              & $       0.318   \pm  0.009      $ & $       0.307   \pm  0.023      $  \\
Quadratic LD coefficient star A           &              0.31 (fixed)         &              0.31 (fixed)          \\
Quadratic LD coefficient star B           &              0.29 (fixed)         &              0.29 (fixed)          \\
$e\cos\omega$                             &              0.0  (fixed)         & $     0.000035  \pm  0.000048   $  \\
$e\sin\omega$                             &              0.0  (fixed)         & $      -0.0010  \pm  0.0017     $  \\
Velocity amplitude star A ($\!$\kms)      & $       70.14   \pm  0.34       $ & $       70.15   \pm  0.37       $  \\
Velocity amplitude star B ($\!$\kms)      & $       82.49   \pm  0.56       $ & $       82.50   \pm  0.58       $  \\
Systemic velocity star A ($\!$\kms)       & $      -29.19   \pm  0.14       $ & $      -29.19   \pm  0.28       $  \\
Systemic velocity star B ($\!$\kms)       & $      -29.30   \pm  0.21       $ & $      -29.30   \pm  0.28       $  \\[3pt]
{\it Derived parameters:} \\
Fractional radius of star~A               & $       0.07685 \pm  0.00053    $ & $       0.07655 \pm  0.00091    $  \\
Fractional radius of star~B               & $       0.05561 \pm  0.00085    $ & $       0.0560  \pm  0.0014     $  \\
Orbital eccentricity                      &              0.0  (fixed)         & $       0.0009  \pm  0.0012     $  \\
Argument of periastron (\degr)            &                n/a                & $    272        \pm  92         $  \\
Light ratio                               & $       0.381   \pm  0.028      $ & $       0.388   \pm  0.028      $  \\[3pt]
\end{tabular}
\end{table}

\begin{figure}[t] \centering \includegraphics[width=\textwidth]{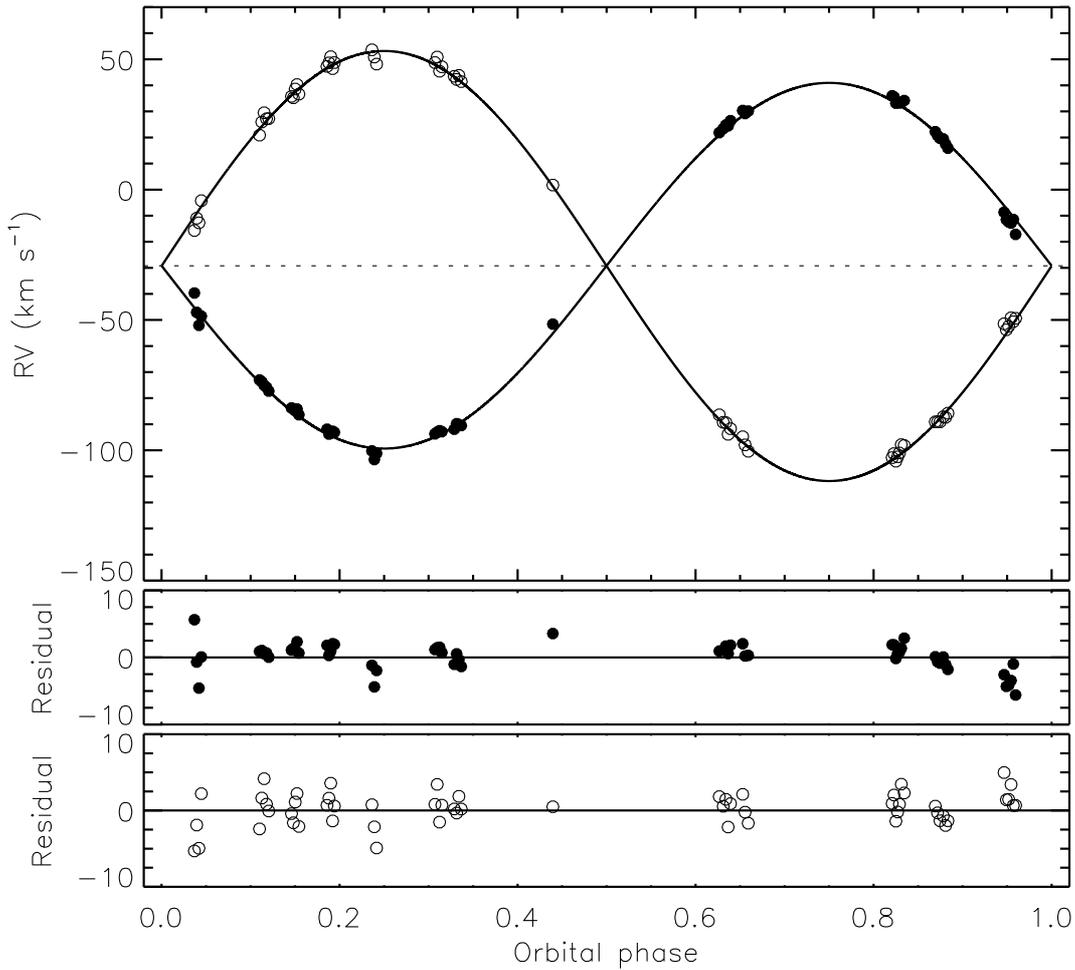} \\
\caption{\label{fig:rv} The RVs of \targ\ from W21 for star~A (filled circles) and star~B
(open circles) plotted versus orbital phase. The solid curves give the best-fitting circular
orbits, which are visually indistinguishable from the best-fitting eccentric orbits. The lower
panels show the residuals of the fits for the two stars individually.} \end{figure}

The uncertainties in the fitted and derived parameters were estimated using Monte Carlo and residual-permutation simulations \cite{Me++04mn2,Me08mn}. The Monte Carlo algorithm requires the data errors to be of a correct size, which was achieved by the scaling to force $\chir=1$. The residual-permutation algorithm successively shifts the residuals of the best fit along the data strings before refitting, so is by design sensitive to red noise and eclipse depth variations. The larger of the two errorbars was adopted for each parameter. Table~\ref{tab:jktebop} gives the best-fitting parameters and their uncertainties for modelling runs assuming a circular or an eccentric orbit. Fig.~\ref{fig:rv} shows the RVs and the fitted circular orbits. The parameter values for the circular- and eccentric-orbit cases are almost identical, and their errorbars are similar, so we adopt the circular orbit as our final result. Inspection of the Monte Carlo and residual-permutation results shows that there is a very strong correlation between the ratio of the radii and the light ratio of the system, as is normally seen in cases where eclipses are shallow and partial (e.g.\ V455~Aur \cite{Me21obs4}). The radius measurements could be improved by obtaining a spectroscopic light ratio, although this will need a large telescope given the relative faintness of \targ.

\section*{How much numerical integration is needed?}

The \kepler\ long-cadence light curve of \targ\ is sampled at a cadence of $t_{\rm samp} = 1765.5$~s (Borucki \cite{Borucki16rpph}) so each eclipse of this EB is covered by only nine or ten datapoints. It is therefore important to account for this when fitting the data, by numerically integrating the model light curve. The method implemented in {\sc jktebop} is to divide each datapoint up into $n_{\rm samp}$ time intervals of equal duration $t_{\rm dur} = t_{\rm samp}/n_{\rm samp}$, calculate the model light curve at the midpoint of each time interval, and then take the mean of the $n_{\rm samp}$ fluxes as the predicted value for that datapoint \cite{Me11mn}. An obvious question is: what is a suitable value of $n_{\rm samp}$? Smaller values risk undersampling the light curve and biasing the best-fitting parameters, whereas larger values take proportionally more computing time. A value of $n_{\rm samp} = 9$ was used in the {\sc jktebop} analysis above.

\begin{figure}[t] \centering \includegraphics[width=\textwidth]{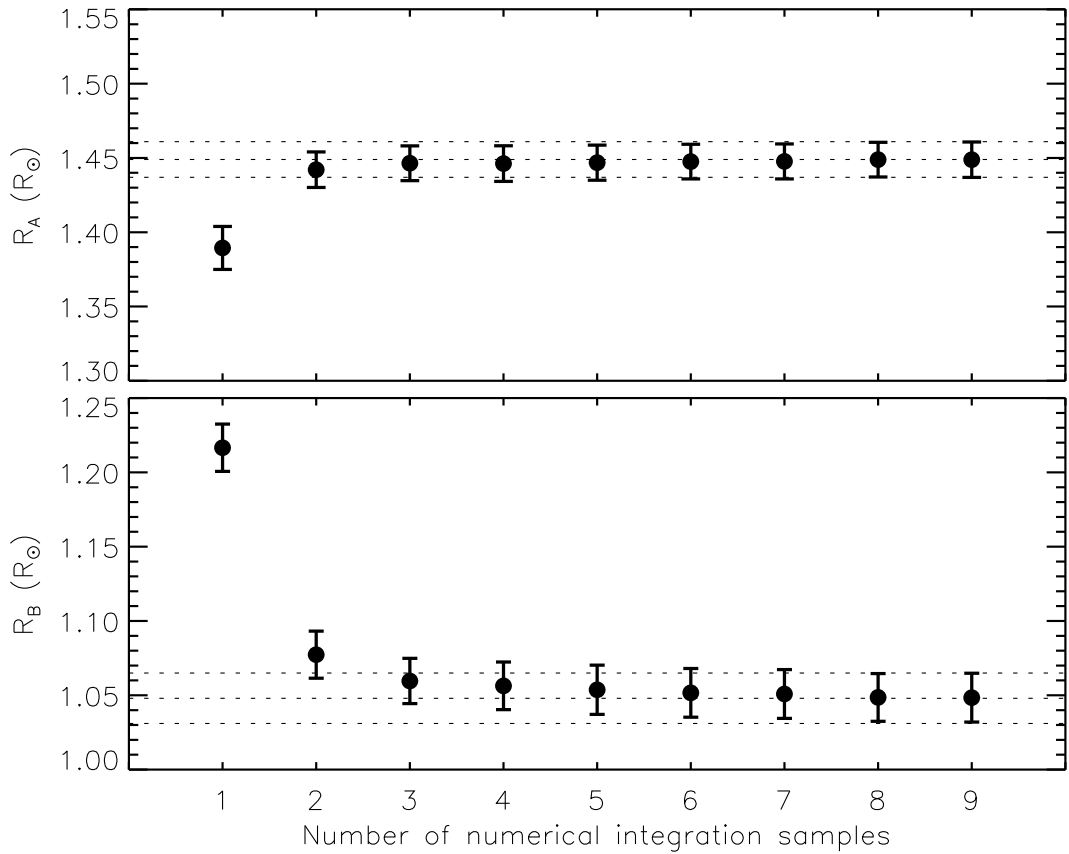} \\
\caption{\label{fig:numint} Best-fitting radii of the stars (star~A in the top
panel and star~B in the bottom panel) with varying amounts of numerical integration.
Errorbars from Monte Carlo simulations are shown. The values and uncertainties
found in the current work are indicated with dotted lines.} \end{figure}

We therefore ran a set of solutions of the \kepler\ light curve (eclipses only) with $n_{\rm samp}$ running from one (equivalent to no numerical integration) to nine (the value used for the main analysis). The results are shown in Fig.~\ref{fig:numint} for the radii of the stars, in the expectation that these are the properties whose measurement is most affected by the sampling rate of the observations. It can be seen that ignoring numerical integration leads to results that are very biased: $R_{\rm A}$ is too small by 4.1$\sigma$ and $R_{\rm B}$ is too large by 10.6$\sigma$. This arises because the smearing of the eclipse shape is best matched using a larger $k$ and a slightly smaller $i$. Using $n_{\rm samp}=2$ immediately leads to a big improvement in the results, and $n_{\rm samp}=3$ is sufficient to bring both radius measurements within the errorbars of the final result. Larger values of $n_{\rm samp}$, perhaps 5 or 8 depending on one's science goals, are needed to measure reliable parameter values. However, the errorbars are reliably measured for much coarser sampling: $n_{\rm samp} \geqslant 2$. It is therefore acceptable to measure parameter values with a larger $n_{\rm samp}$, but then run the error analysis algorithms with a lower $n_{\rm samp}$ to save computing time.

These conclusions are specific to \targ\ but are likely to hold for most EBs with eclipses of similar duration. They are comparable to what Southworth \cite{Me11mn} found for the synthetic light curve of a transiting extrasolar planetary system resembling Kepler-6. Similar conclusions were also reached by Kipping \cite{Kipping10mn}. The most important point is that if the data have been temporally averaged then the model one fits to it must be treated in the same way.


\section*{Physical properties of \targ}

To determine the physical properties of the \targ\ system we used the values and uncertainties of $r_{\rm A}$, $r_{\rm B}$, $P$, $i$,  $K_{\rm A}$ and $K_{\rm B}$ from Table~\ref{tab:jktebop}. The calculations were performed with the {\sc jktabsdim} code \cite{Me++05aa}, which calculates the physical properties using standard formulae and propagates uncertainties via perturbation. We adopted the \Teff\ values of the stars from W21, but increased the errorbar for star~B because the quoted value is measured relative to the \Teff\ of star~A so therefore should include its uncertainty. The ratio of the \Teff s is consistent with the surface brightness ratio measured from the light curve (Table~\ref{tab:jktebop}). The results are given in Table~\ref{tab:absdim}.

To determine the distance to the system we used its apparent magnitudes in $B$ and $V$ from APASS DR9 and in $H$ and $K$ from 2MASS (see Table~\ref{tab:info}); the 2MASS $J$ magnitude is unreliable \cite{Cutri+03book}. The interstellar extinction was estimated as $E(B-V) = 0.07 \pm 0.03$\,mag obtained using the {\sc stilism}\footnote{\texttt{https://stilism.obspm.fr}} online tool (Lallement \etal\ \cite{Lallement+14aa,Lallement+18aa}). Its distance of $1534 \pm 27$~pc was calculated using the surface brightness method from Southworth \etal\ \cite{Me++05aa} and the $K$-band surface brightness calibrations from Kervella \etal\ \cite{Kervella+04aa}. This is significantly smaller than the distance of $1626 \pm 36$~pc from the parallax of the system in \gaia\ EDR3 \cite{Gaia21aa}.

We are unable to deduce the reason why our distance is discrepant with that from \gaia. The 2MASS apparent magnitudes are single-epoch so may have been obtained at a time of particularly strong starspot activity -- we have no data available to provide an independent cross-check but the required change of approximately $+0.1$\,mag in $K$ is much larger than the brightness modulation due to starspot activity seen in Fig.~\ref{fig:time}. The 2MASS observations were taken at phase $0.9801 \pm 0.0001$ based on our ephemeris in Table~\ref{tab:jktebop}, which is close to but confidently outside primary eclipse (see Fig.~\ref{fig:ecl}). The APASS $BV$ magnitudes come from six epochs so are very unlikely to be made significantly fainter by eclipses. A larger set of \Teff\ values allied with stronger reddening also requires implausible changes -- increases of 1300~K in both \Teff s and 0.23~mag in $E(B-V)$ -- to fix the discrepancy. Stellar radii larger by 5\% would also be sufficient, but far beyond our 1$\sigma$ errorbars in Table~\ref{tab:absdim} and significantly greater than those found by W21. We leave this mystery for future study.

\begin{table} \centering
\caption{\em Physical properties of \targ\ defined using the nominal solar units given by
IAU 2015 Resolution B3 (Ref.\ \cite{Prsa+16aj}). The \Teff s are from W21. \label{tab:absdim}}
\begin{tabular}{lr@{\,$\pm$\,}lr@{\,$\pm$\,}l}
{\em Parameter}        & \multicolumn{2}{c}{\em Star A} & \multicolumn{2}{c}{\em Star B}    \\[3pt]
Mass ratio                                  & \multicolumn{4}{c}{$0.8503 \pm 0.0071$}       \\
Semimajor axis of relative orbit (\Rsunnom) & \multicolumn{4}{c}{$18.854 \pm 0.081$}        \\
Mass (\Msunnom)                             &  1.252  & 0.018       &  1.065  & 0.013       \\
Radius (\Rsunnom)                           &  1.449  & 0.012       &  1.048  & 0.017       \\
Surface gravity ($\log$[cgs])               &  4.214  & 0.007       &  4.424  & 0.013       \\
Density ($\!\!$\rhosun)                     &  0.412  & 0.009       &  0.924  & 0.043       \\
Synchronous rotational velocity ($\!\!$\kms)&  11.76  & 0.10        &  8.51   & 0.14        \\
Effective temperature (K)                   &    6500 & 50          &    5980 & 70          \\
Luminosity $\log(L/\Lsunnom)$               &   0.555 & 0.014       &   0.103 & 0.025       \\
$M_{\rm bol}$ (mag)                         &   3.353 & 0.037       &   4.483 & 0.061       \\
Distance (pc)                               & \multicolumn{4}{c}{$1534 \pm 27$}             \\[3pt]
\end{tabular}
\end{table}


\section*{Summary and conclusions}

\targ\ is a dEB containing two F-type stars that both show brightness variations due to starspots. We have extracted the data around eclipse from the light curve of this system obtained using the \kepler\ satellite and fitted them with {\sc jktebop} to determine the photometric properties of the system. Numerical integration was used to account for the low sampling rate of the \kepler\ long-cadence data. We included published RVs for the two stars to determine their masses and radii. With the inclusion of spectroscopic measurements of the \Teff s of the two stars, we determined their luminosities and the distance to the system. The distance measurement is slightly shorter than that from \gaia\ EDR3 -- at the level of 2.0$\sigma$ -- and we have not found a good explanation for this minor discrepancy.

We investigated the amount of numerical integration needed to fit the \kepler\ data for \targ. We found that ignoring the sampling rate leads to radius measurements that are wrong by $-4\sigma$ (for $R_{\rm A}$) and $+10\sigma$ (for $R_{\rm B}$). These biases are fixed by the application of only a small amount of numerical integration to the model light curve during the fitting process: merely doubling the model sampling rate is sufficient for approximate solutions but more precision is needed for final results.

We can now turn to the original prompt of this work: the analysis of W21 presented very different errorbars for circular- and eccentric-orbit solutions; no uncertainty was given for the mass of star~B; and there was no mention of the sampling cadence. For the first point, we find very similar results for the two options and adopted the circular-orbit results for our own calculations. For the second, the uncertainty in the mass of star~B is 1.2\%. For the third, the results of W21 are much closer to our own when we used numerical integration so we conclude that they did account for this effect in their analysis.

A wider comparison between the results of W21 (adopting the circular-orbit values) and our own shows good agreement for the radius of star~B: $1.048 \pm 0.017$\Rsun\ (this work) versus $1.05 \pm 0.03$\Rsun\ (W21); but not star~A: $1.449 \pm 0.012$\Rsun\ (this work) versus $1.52 \pm 0.03$\Rsun\ (W21). The reason is not clear but could be related to the different ways in which the spot activity was accounted for. Our mass measurements, though, are in unexpectedly poor accord: $1.252 \pm 0.018$\Msun\ versus $1.32 \pm 0.02$\Msun\ for star~A, and $1.065 \pm 0.013$\Msun\ versus 1.12\Msun\ (W21, no errorbar quoted) for star~B. As the mass measurements are primarily dependent on the RVs, and both studies used the same ones, it is not clear how this discrepancy could have occurred. Our results have come from extensively-tested codes and analysis methods \cite{Me13aa,Maxted+20mn,Me21obs5} so should be reliable.

The final motivation for the current study was to see if the available data were sufficient to establish the masses and radii of the component stars of \targ\ to 2\% or better. Table~\ref{tab:absdim} shows that it was indeed possible. This dEB has now been added to the Detached Eclipsing Binary Catalogue (\debcat\footnote{\texttt{https://www.astro.keele.ac.uk/jkt/debcat/}}, Ref. \cite{Me15debcat}) and can in future be used to help calibrate theoretical models of stars with masses close to that of our Sun.


\section*{Acknowledgements}

This paper includes data collected by the \kepler\ mission and obtained from the MAST data archive at the Space Telescope Science Institute (STScI). Funding for the \kepler\ mission is provided by the NASA Science Mission Directorate. STScI is operated by the Association of Universities for Research in Astronomy, Inc., under NASA contract NAS 5–26555.
The following resources were used in the course of this work: the NASA Astrophysics Data System; the SIMBAD database operated at CDS, Strasbourg, France; and the ar$\chi$iv scientific paper preprint service operated by Cornell University.


\bibliographystyle{obsmaga}

\end{document}